\documentclass[]{aa}

\usepackage{amsmath}
\usepackage{txfonts}
\usepackage{natbib}
\usepackage{graphicx}

\newcommand{\elie}{{\mathbf E}}
\newcommand{\uin}{u_{\rm in}}
\newcommand{\vout}{v_{\rm out}}
\newcommand{\ain}{a_{\rm in}}
\newcommand{\aout}{a_{\rm out}}
\newcommand{\elik}{{\mathbf K}}

\begin{document}

\title{Generation of potential/surface density pairs in flat disks}
\subtitle{Power law distributions}
\titlerunning{Potential/surface density pairs in flat disks}

\author{Jean-Marc Hur\'e\inst{1,}\inst{2}, Didier Pelat\inst{2} and Arnaud Pierens\inst{3}}

\offprints{Jean-Marc Hur\'e}

\institute{Universit\'e Bordeaux 1/CNRS/OASU/UMR 5804/LAB, 2 rue de l'Observatoire, BP 89, 33270 Floirac, France\\
\email{jean-marc.hure@obs.u-bordeaux1.fr}
\and
LUTh/Observatoire de Paris-Meudon-Nancay, Place Jules Janssen, 92195 Meudon Cedex, France\\
\email{didier.pelat@obspm.fr}
\and
Astronomy Unit, Queen Mary, University of London, Mile end Road, London E1 4NS, UK\\
\email{a.pierens@qmul.ac.uk}
}

\date{Received ??? / Accepted ???}

\abstract
{}
{We report a simple method to generate potential/surface density pairs in flat axially symmetric finite size disks.}
{Potential/surface density pairs consist of a ``homogeneous'' pair (a closed form expression) corresponding to a uniform disk, and a ``residual'' pair. This residual component is converted into an infinite series of integrals over the radial extent of the disk. For a certain class of surface density distributions (like power laws of the radius), this series is fully analytical.}
{The extraction of the homogeneous pair is equivalent to a convergence acceleration technique, in a matematical sense. In the case of power law distributions (i.e. surface densities of the form $\Sigma(R) \propto R^s$), the convergence rate of the residual series is shown to be cubic inside the source. As a consequence, very accurate potential values are obtained by low order truncation of the series. At zero order, relative errors on potential values do not exceed a few percent typically, and scale with the order $N$ of truncation as $1/N^3$. This method is superior to the classical multipole expansion whose very slow convergence is often critical for most practical applications. }
{}
{}


\maketitle

\section{Introduction}

The construction of potential/density pairs is a common concern in the context of galactic dynamics \cite[e.g.][]{binneytremaine87}. More generally, the accurate determination of the gravitational potential corres\-ponding to a given mass distribution is a critical step in astrophysical disk models and numerical simulations.  Unfortunately, potential/density pairs are not closed-form expressions, except in a few cases only \cite[][]{mestel63,dezeewpfenniger88, binneytremaine87}, and most known pairs involve infinite matter distribution \cite[e.g.][]{qian93,cuddeford93,earn96,binneytremaine87}, whereas astrophysical disks are finite (in size and mass).

For flat (i.e. zero thickness) and finite disks of interest here, the potential is mainly accessible through the double integration of the surface density weighted by the Green function $1/|\vec{r}-\vec{r}'|$ \cite[see][for a review]{binneytremaine87}. This integral approach is very tricky due to the inevitable occurrence of singularities everywhere inside matter, i.e. when $\vec{r} \rightarrow \vec{r}'$. The difficulty is circumventet when the Green function is converted into series \cite[][]{kellogg29}. However, these are infinite and alternate series which do not converge inside sources, although their integration over the distribution does \cite[][]{durand64}. The efficiency of the series representation is therefore strongly misleading in practice because series truncations do not lead to a numerically stable description of the potential inside disks \cite[][]{clement74}. The errors in potential values can be as large as a few percent on average, even when the order of truncation attain several tens \cite[][]{stonenorman92}. Singularities are no more present when the potential is expressed in terms of integrals of Bessel functions. But the oscillatory behavior of Bessel functions as well as their infinite definition domain pose technical difficulties, even for infinitely extended disks. Potential/density pairs can also be constructed by superposition of homeoids conveniently shrunk along one axis \cite[][]{binneytremaine87}, or by parametric expansion and ordering of the associated Poisson equation \cite[][]{ciottibertin05}.

\citet[][]{hurepierens05} have reported a numerical method to compute the gravitational potential in flat axially symmetric disks with great accuracy by a correct treatment of the singularity. When the surface density profile is conveniently split into two components (namely, a homogeneous component plus a residual one), the contribution of the singularity to the potential can be accounted for exactly through a closed-form expression. In this paper, we explore this ``splitting method'' from a purely theoretical point of view in search for fully analytical solutions (i.e. potential/surface density pairs). In particular, it is shown that the residual component is an infinite series of radial integrals which are known analytically for a certain class of surface density distributions. This is the case of power law profiles which are of great astrophysical interest \citep[e.g.][]{pringle81}. The convergence rate of the residual component is then cubic. This is a major issue, when compared with the classical multipole expansion whose converge speed is prohibitively low inside sources.

This paper is organized as follows. In Sect. \ref{sec:intro}, we briefly recall the density splitting method for one dimensional disks \citep{hurepierens05}. We expand the residual component as a infinite series of the integrals over the modulus of the complete elliptic integral of the first kind. We then show that the construction of potential/surface density pairs is possible for a certain class of surface density distribution. The impact of the series truncation on potential values is discussed in Sect. \ref{sec:trunc}. The case of surface densities scaling as a power of the radius is investigated in detail in Sect. \ref{sec:plaws} (a Fortran 90 code is available on request). For this kind of distribution, the convergence rate is cubic. This is established in Sect. \ref{sec:crate}. The paper ends with a summary and concluding remarks.

\begin{figure}[h]
\includegraphics[width=8.9cm]{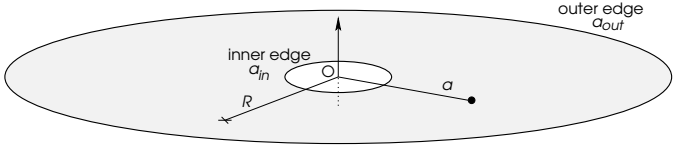}
\caption{A finite axially symmetric disk.}
\label{fig:scheme.xfig.eps}
\end{figure}

\section{Exact potential/surface density pairs in flat axisymmetric disks}
\label{sec:intro}

\subsection{Outline of the density splitting method}

\begin{figure}[h]
\includegraphics[width=8.9cm]{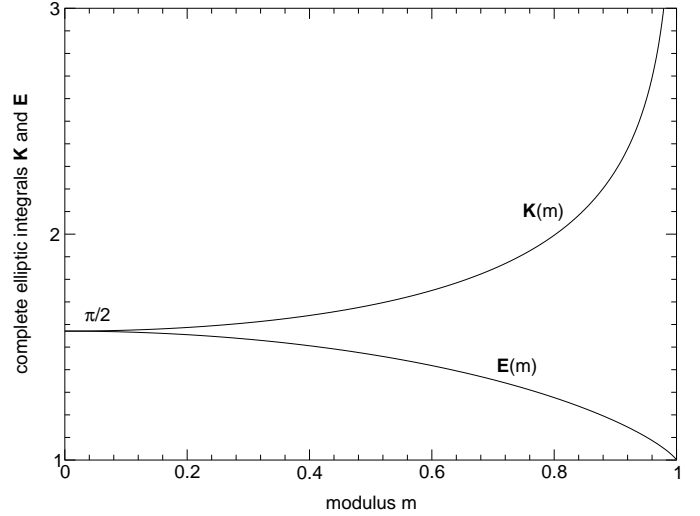}
\caption{Complete elliptic integrals of the first and second kinds versus the modulus $m$.}
\label{fig:ek.eps}
\end{figure}

The gravitational potential $\Psi$ in the plane of a flat axially symmetric disk as pictured in Fig. \ref{fig:scheme.xfig.eps} is given by the exact expression \cite[e.g.][]{durand64}:
\begin{equation}
\Psi(R) = -2G \int_{\ain}^{\aout}{\Sigma(a)\sqrt{\frac{a}{R}}m \elik(m)da},
\label{eq:psi}
\end{equation}
where
\begin{equation}
\elik(m)=\int_0^{\pi/2}{\frac{d \phi}{\sqrt{1-m^2 \sin^2 \phi}}}
\end{equation}
is the complete elliptic integral of the first kind,
\begin{equation}
m=\frac{2\sqrt{aR}}{a+R}
\end{equation}
is the modulus of $\elik$ (with $0 \le m \le 1$), $a$ and $R$ are polar cylindrical radii ($a$ refering to the matter distribution), $\ain \ge 0$ is the radius of the inner edge, $\aout > \ain$ is the radius of the outer edge, $\Sigma(a)$ is the surface density, and $G$ is the Gravitation constant. The function $\elik$ is plotted versus $m$ in Fig. \ref{fig:ek.eps}; it is characterized by a divergence for $m \rightarrow 1$. This is a logarithmic divergence \cite[see e.g.][]{kellogg29,mestel63}, since we have \cite[][]{gradryz65}:
\begin{equation}
\lim_{m \rightarrow 1}\elik(m) = \ln \frac{4}{\sqrt{1-m^2}}\left[ 1 + {\cal O}(1-m^2) \right].
\end{equation}
Despite this singularity which occurs everywhere inside the disk (i.e. for $\ain \le R \le \aout$), the potential is generally finite everywhere.

In order to determine $\Psi$ inside the disk, we set:
\begin{equation}
u=\frac{a}{R} \le 1 \qquad {\rm and} \qquad v = \frac{R}{a} \le 1,
\end{equation}
and so, $u$ locates matter of the inner disk $[\ain,R]$, whereas $v$ is for the outer part $[R,\aout]$. Using the transformation \citep[e.g.][]{gradryz65}:
\begin{equation}
\elik\left(\frac{2\sqrt{x}}{1+x}\right) = (1+x) \elik(x),
\end{equation}
where $0 \le x \le 1$, Eq.(\ref{eq:psi}) reads:
\begin{equation}
\Psi(R) = -4GR \left[ \int_{\uin}^1{\Sigma(u) \elik(u)udu} - \int_1^{\vout}{\Sigma(v) \frac{\elik(v)}{v^2}dv}\right],
\label{eq:psibis}
\end{equation}
where $\uin = \ain/R \le 1$ and $\vout = R/\aout \le 1$ correspond to the position of the disk edges. As quoted, $\Sigma$ must now be regarded as a function of $u$ in the first integral, and as a function of $v=1/u$ in the second one. According to the ``density splitting'' method \citep{hurepierens05}, we divide $\Sigma$ into two components, namely:
\begin{equation}
\Sigma(a) = \Sigma_0+\delta \Sigma(a,R),
\label{eq:sigma}
\end{equation}
where $\Sigma_0 \equiv \Sigma(R)$ is the value of the surface density at the radius of the singularity (where the moduli $m$, $u$ and $v$ are unity), and $\delta \Sigma$ is the ``residual'' surface density profile. The potential then takes the form:
\begin{equation}
\Psi(R) = \Psi_0(R) + \delta \Psi(R),
\label{eq:psitot}
\end{equation}
where
\begin{equation}
\Psi_0(R) =  -4GR \Sigma_0 \left[ \int_{\uin}^1{\elik(u)udu} - \int_1^{\vout}{\frac{\elik(v)}{v^2}dv} \right]
\label{eq:psi2}
\end{equation}
is the potential due to a radially homogeneous disk with constant surface density $\Sigma_0$, and
\begin{flalign}
\nonumber
\delta \Psi(R) & =  - 4GR \left[ \int_{\uin}^1{\delta \Sigma(u)\elik(u)udu} \right.\\
& \qquad  \qquad  \qquad  \qquad     \left. - \int_1^{\vout}{\delta \Sigma(v)\frac{\elik(v)}{v^2}dv} \right]
\label{eq:deltapsi}
\end{flalign}
is the ``residual'' potential associated with $\delta \Sigma$. Note that $\Sigma_0$ varies from place to place in the disk, as well as $\delta \Sigma$. As demonstrated elsewhere \citep[see also][]{hure05}, this technique of density splitting can be numerically very accurate, for two reasons. First, $\Psi_0$ is known in a closed form:
\begin{equation}
\Psi_0(R) = -4GR \Sigma_0 \left[ \frac{\elie(\vout)}{\vout} -\elie(\uin) + {\uin'}^2 \elik(\uin) \right],
\label{eq:psi0}
\end{equation}
and is finite everywhere in the disk mid-plane. In Eq.(\ref{eq:psi0}), ${\uin'}~=~\sqrt{1-\uin^2}$ is the complementary modulus and $\elie$ is the complete elliptic integral of the second kind (see Fig. \ref{fig:ek.eps}):
\begin{equation}
\elie(m)=\int_0^{\pi/2}{d\phi \sqrt{1-m^2 \sin^2 \phi}}.
\end{equation}
Second, each integrand in Eq.(\ref{eq:deltapsi}) now vanishes at the singularity (this is the role of the splitting), making the two integrals easily computable with standard quadrature rules. 

\subsection{The residual potential from series representation}

In order to determine analytical expressions for $\Psi$ on the basis of the splitting method, we consider the expansion of the elliptic integral of the first kind over its modulus \cite[e.g.][]{gradryz65}, namely:
\begin{equation}
\left\{
\begin{aligned}
&\elik(m)= \frac{\pi}{2}  \sum_{n=0}^{\infty}{\gamma_n m^{2n}},\\
&\gamma_0 =1,\\
&\gamma_n = \left[\frac{(2n-1)!!}{2^n \, n!}\right]^2 = \gamma_{n-1} \left(\frac{2n-1}{2n}\right)^2, \qquad n \ge 1.
\label{eq:kseries}
\end{aligned}
\right.
\end{equation}
When this infinite series is inserted into Eq.(\ref{eq:deltapsi}), the residual potential $\delta \Psi$ becomes an infinite series. Let
\begin{equation}
S_N(x_n) = \sum_{n=0}^N{x_n},
\label{eq:sn}
\end{equation}
be the sum of $N+1$ terms $x_0, x_1,\dots, x_N$. The residual potential is then given by:
\begin{equation}
\delta \Psi = S_\infty \left(\delta \Psi_n \right),
\label{eq:psiinf}
\end{equation}
with for any  $n \ge 0$
\begin{equation}
\begin{aligned}
\delta \Psi_n   =  - 2 \pi G R \gamma_n & \left[ \int_{\uin}^1{\delta \Sigma(u)u^{2n+1}du} \right.\\
 & \qquad \qquad \left. - \int_1^{\vout}{\delta \Sigma(v)v^{2n-2}dv} \right].
\label{eq:deltapsin}
\end{aligned}
\end{equation}
Note that this expression for $\delta \Psi$ is still {\it exact}, provided all the terms $\delta \Psi_n$ of the series are included. It follows that potential/surface density pairs $(\Psi_0 + \delta \Psi,\Sigma_0+\delta \Sigma)$ can easily be constructed for flat disks with finite size (or not, depending on $\uin$ and $\vout$). These pairs are analytical if the residual pair ($\delta \Psi, \delta \Sigma)$ is analytical, that is, if the product of $\Sigma(a)$ by a power law of $a$ is analytically integrable. We immediately see that many simple functions $\Sigma(a)$ are concerned. In particular, if $\Sigma$ is a power law of the radius $a$ |as often met in astrophysical disk models|, then $\delta \Psi_n$ is a power law of the radius too (more precisely, a mixture of four power laws; see Sect. \ref{sec:plaws}). Obviously, pairs determined through Eqs.(\ref{eq:psiinf}) and (\ref{eq:deltapsin}) should generally remain in the form of an infinite series with no equivalent closed form expression. This is not a big problem, first because working with infinite expansions is common in potential theory, and second because the residual potential $\delta \Psi$ is a rapidly converging series (see Sect. \ref{sec:crate}). 

\section{Approximate potential from truncated series}
\label{sec:trunc}

\subsection{General remark}

\begin{figure}
\includegraphics[width=8.9cm]{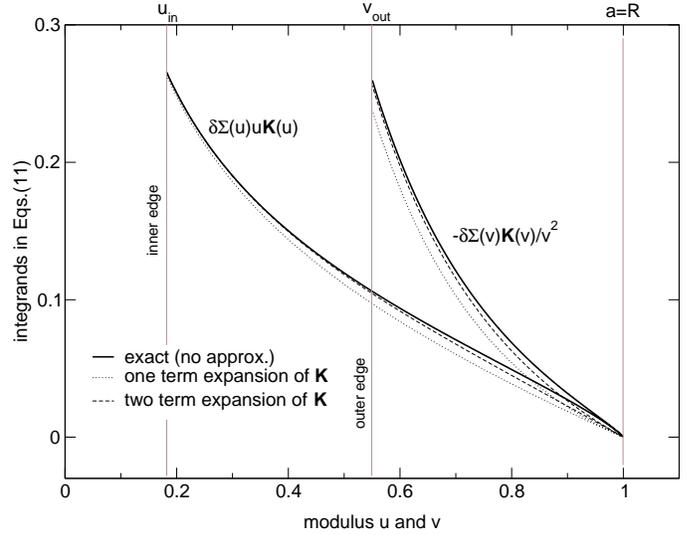}
\caption{Typical shape of the integrands appearing in Eq.(\ref{eq:deltapsi}) computed from different approximations for the function $\elik$. In this example, the disk has inner edge $\ain=0.1$ and outer edge $\aout=1$, the surface density obeys a power law of the radius with $\Sigma(a) \propto a^{-3/2}$ (that is $s=-1.5$), and the radius is $R=\frac{1}{2}(\ain+\aout)$.}
\label{fig:mpa2.eps}
\end{figure}

The estimate of $\Psi_0$ should generate no significant error since special functions $\elie$ and $\elik$ can be determined within the computer precision from a numerical library. In contrast, the computation of $\delta \Psi$ is not errorless because, in practice, any series is necessarily truncated at a certain order $N$. Does the truncation produce large errors in potential values?  In order to answer this question, it is important to stress again that the occurrence of the condition $\delta \Sigma = 0$ enables the neutralization of the logarithmic singularity,  making the remaining integrands fully regular. As a consequence, the value of the residual potential $\delta \Psi$ is {\it mainly determined by the shape of the integrands far from the singularity}, that is {\it close to the edges}. In other words, neither the precise behavior of the complete elliptic integral $\elik$ nor the shape of the residual surface density profile $\delta \Sigma$ {\it around the singularity} are critical to estimate $\delta \Psi$ with a certain accuracy. We therefore expect that $\delta \Psi$ can be accurately determined if $\elik$ is replaced by any approximation which preserves at best the values of this function around the bounds $\uin$ and $\vout$ rather than at the radius of the singularity (where $m=u=v=1$). From this point of view, the series defined by Eq.(\ref{eq:kseries}) is fully appropriate as it converges very rapidly for moduli significantly less than unity. Figure \ref{fig:mpa2.eps} shows a typical example where $\Sigma$ is a power law of the radius. We clearly see that the two integrands $\delta \Sigma(u) u \elik(u)$ and $\delta \Sigma(v)\elik(v)/v^2$ appearing in Eq.(\ref{eq:deltapsi}) are quite insensitive to the approximation adopted for $\elik$ (one-term, two-term expansions or exact). This conclusion is independent of the $\Sigma$-profile and of $R$ as well.

\subsection{The first terms}

For a truncation of the series at order $N$, the potential is given by the approximate value:
\begin{equation}
\Psi_{\rm approx.} \equiv \Psi_0+S_N(\delta \Psi_n) \ne \Psi,
\label{eq:psiapprox}
\end{equation}
and we have asymptotically:
\begin{equation}
 \Psi_0 + \lim_{N \rightarrow \infty}{\left[ S_N ( \delta \Psi_n ) \right]} = \Psi.
\label{eq:psiasympt}
\end{equation}
With only one term in Eq.(\ref{eq:kseries}), we have $S_0(\delta \Psi_n) = \delta \Psi_0$ with:
\begin{equation}
\label{eq:deltapsi0}
\delta \Psi_0 =  -2 \pi GR \left[ \int_{\uin}^1{\delta \Sigma(u)u du} - \int_1^{\vout}{\delta \Sigma(v)\frac{dv}{v^2}} \right],
\end{equation}
and so the approximate potential is $\Psi_{\rm approx.} = \Psi_0 + \delta \Psi_0$. For a two-term expansion of the complete elliptic integral of the first kind, the residual potential is $S_1(\delta \Psi_n) = \delta \Psi_0+\delta \Psi_1$, where $\delta \Psi_0$ is given by Eq.(\ref{eq:deltapsi0}), and:
\begin{equation}
\delta \Psi_1  =  -\frac{\pi}{2} G R \left[ \int_{\uin}^1{\delta \Sigma(u)u^3du} - \int_1^{\vout}{\delta \Sigma(v)dv} \right].
\label{eq:deltapsi1}
\end{equation}
The approximate potential is then $\Psi_{\rm approx.} = \Psi_0+\delta \Psi_0 +\delta \Psi_1$, and so on. Each new term added into the expansion brings a new component $\delta \Psi_n$ defined by Eq.(\ref{eq:deltapsin}). We then expect that the more terms in the expansion, the more accurate the potential. If the series converges rapidly (this is the case of  power law distributions; see below), the absolute error made on $\Psi$ is roughly $|\delta \Psi_{N+1}|$.

\section{The case of power-law surface density profiles}
\label{sec:plaws}

Let us consider the case where $\Sigma$ is a power-law of the radius, namely:
\begin{equation}
\Sigma = \sigma_0 \left(\frac{a}{\aout} \right)^s
\label{eq:sigmaplaw}
\end{equation}
where $\sigma_0$ is the surface density at the outer edge of the disk. Power-law surface density solutions are found in many disk models \cite[e.g.][]{pringle81} and match observations \cite[e.g.][]{guiloteaudutrey98}, with $s<0$ in general. We then have:
\begin{equation}
\Sigma_0 = \sigma_0 \left(\frac{R}{\aout} \right)^s
\end{equation}
and so:
\begin{equation}
\left\{
\begin{aligned}
&\delta \Sigma = \Sigma_0  \left( u^s -1 \right)  \quad \text{for} \quad u \le 1,\\
&\delta \Sigma = \Sigma_0  \left( {v^{-s}}-1\right)  \quad \text{for} \quad v \le 1.
\end{aligned}
\right.
\end{equation}

\begin{figure}
\includegraphics[width=8.9cm]{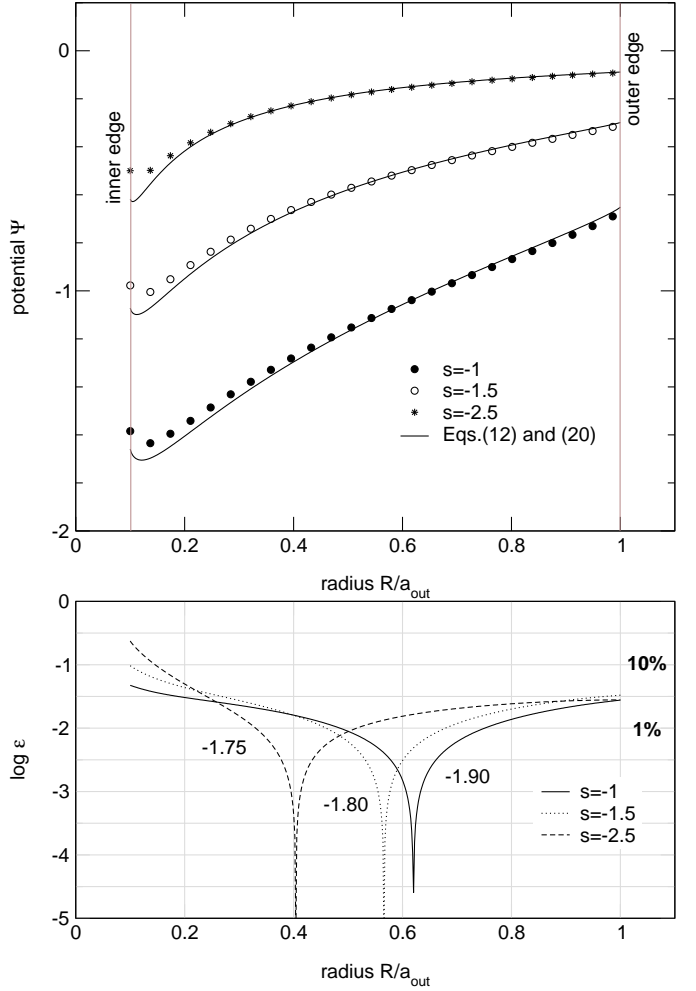}
\caption{{\it Top}: potential ({\it lines}) inside the disk for three different surface density exponents $s$  when computed with a one-term expanded elliptic integral (i.e. $\delta \Psi \equiv \delta \Psi_0$). Reference values $\Psi_{\rm ref.}$ ({\it symbols}) are obtained numerically within the computer precision. {\it Bottom} decimal logarithm of the relative error (the average of $\epsilon$ is also indicated).}
\label{fig:mpapl.eps}
\end{figure}

\subsection{Order zero}

For a one-term expansion of the complete elliptic integral of the first kind, the residual potential is given by:
\begin{equation}
\label{eq:deltapsi0plaw}
\delta \Psi_0 =  2\pi G R \Sigma_0  \left[ \int_{\uin}^1{(1-u^s)udu} + \int_1^{\vout}{\left(v^{-s}-1\right)\frac{dv}{v^2}} \right].
\end{equation}

If $s \ne \{-2, -1\}$, this is (see Appendix \ref{sec:sp0} for these two cases):
\begin{equation}
\begin{aligned}
\delta \Psi_0 & =  2  \pi G R \Sigma_0  \left[\frac{\uin^{s+2}-1}{s+2} + \frac{1-\uin^2}{2} \right. \\
            & \qquad \qquad\qquad\qquad\left. +\frac{1-\vout^{-s-1}}{s+1} + \frac{1}{\vout}-1 \right].
\end{aligned}
\end{equation}

Figure \ref{fig:mpapl.eps} displays the approximate potential $\Psi_{\rm approx.} = \Psi_0 + \delta \Psi_0$ for disks with inner edge $\ain=0.1$, outer edge $\aout=1$ and power law exponents $s=-1$, $-1.5$ and $-2.5$. References potential values $\Psi_{\rm ref.}(R)$ have been determined within the computer precision from the numerical splitting method \citep{hurepierens05} and compared to $\Psi_{\rm approx.}$. The decimal logarithm of the relative error $\epsilon \equiv |1- \Psi_{\rm approx.}/\Psi_{\rm ref.}|$ is also shown in Fig. \ref{fig:mpapl.eps}. We see that the potential is reproduced with an accuracy better than $\sim 2 \%$ (on average) in the entire disk with only one term in the expansion of the $\elik$-function. This is already remarkable. The accuracy is relatively uniform, and it is slightly better at the outer edge. This point is particularly interesting to model young stellar objects or Active Galactic Nuclei. In these systems, the inner disk is dynamically under the control of a massive central object (a proto-star or a black hole), and so there is no need for very accurate potential values. On the other hand, in the outer regions where disks can be self-gravitating \cite[e.g.][]{hure00}, a reliable description of the potential is required.

\subsection{First order}

The accuracy of potential values can be improved by adding a second term. For $N~=~1$, we find:
\begin{equation}
\delta \Psi_1 =  \frac{\pi}{2} G R \Sigma_0 \left[ \int_{\uin}^1{(1-u^s)u^3du} + \int_1^{\vout}{\left(v^{-s}-1\right)dv} \right].
\end{equation}

If $s \ne \{-4, 1\}$ (see Appendix \ref{sec:sp1} for these two cases), the expression is:
\begin{equation}
\begin{aligned}
\delta \Psi_1 & =  2  \pi G R \Sigma_0 \left[ \frac{\uin^{s+4}-1}{4(s+4)} + \frac{1-\uin^4}{16} \right. \\
& \qquad \qquad \qquad \qquad \left.  + \frac{1- \vout^{-s+1}}{4(s-1)} + \frac{1 - \vout}{4} \right].
\end{aligned}
\end{equation}

\begin{figure}
\includegraphics[width=8.9cm]{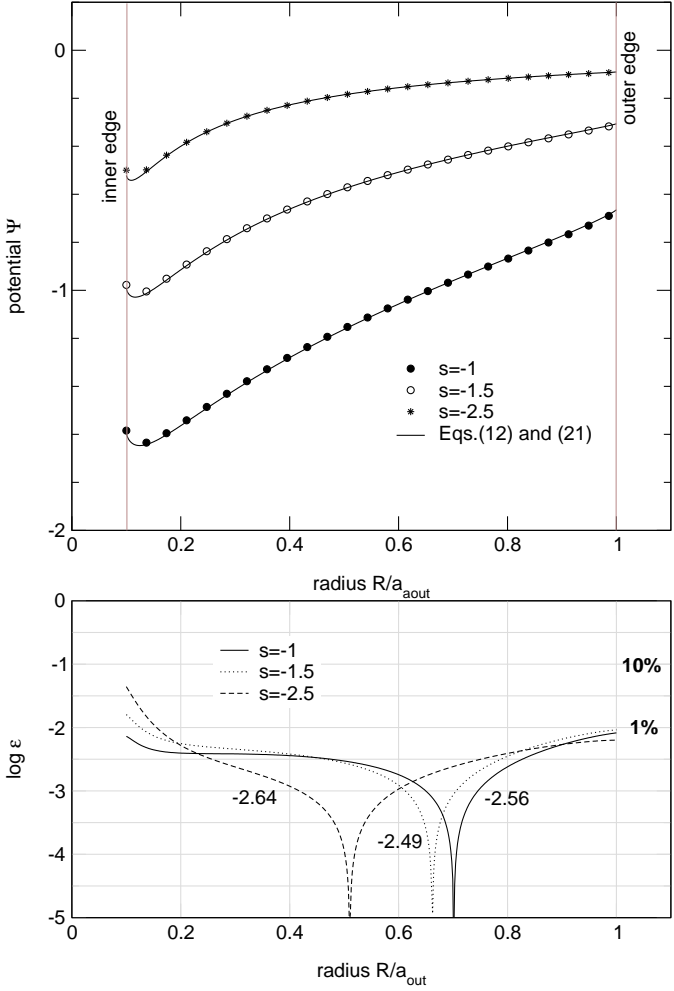}
\caption{Same legend as for Fig. \ref{fig:mpapl.eps} but the residual potential $\delta \Psi$ is determined from a two-term expanion of the complete elliptic integral of the first kind (i.e. $\delta \Psi \equiv \delta \Psi_0+  \delta \Psi_1$).}
\label{fig:mpapl1.eps}
\end{figure}

Figure \ref{fig:mpapl1.eps} displays the results obtained with a two-term expansion of $\elik$, i.e. {\bf $\Psi_{\rm approx.} = \Psi_0 + \delta \Psi_0 + \delta \Psi_1$}, under the same conditions as for Fig. \ref{fig:mpapl.eps}. We see that the relative accuracy of mid-plane potential values (of the order of $0.3\%$ on average) is now better by a factor of $\sim 5$. Such a level of accuracy is probably sufficient for many disk models and applications. However, since the process of adding more and more terms represents no technical difficulty and converges rapidly (see Sect. \ref{sec:crate}), it can be repeated until a given accuracy is reached. In parti\-cular, for $s+2n+2\ne0$ and $s-2n+1 \ne 0$, we have (see the Appendix \ref{sec:spn} for these two cases)
\begin{equation}
\begin{aligned}
\delta \Psi_n&= 2 \pi G \Sigma_0 R \gamma_n \left[\frac{1-\uin^{2(n+1)}}{2(n+1)}-\frac{1- \uin^{s+2(n+1)}}{s+2(n+1)} \right. \\
&\qquad \qquad \qquad \qquad  \left. + \frac{\vout^{2n-s-1}-1}{2n-s-1} - \frac{\vout^{2n-1}-1}{2n-1} \right].
\label{eq:deltapsinplaws}
\end{aligned}
\end{equation}

\begin{figure}[h]
\includegraphics[width=8.9cm]{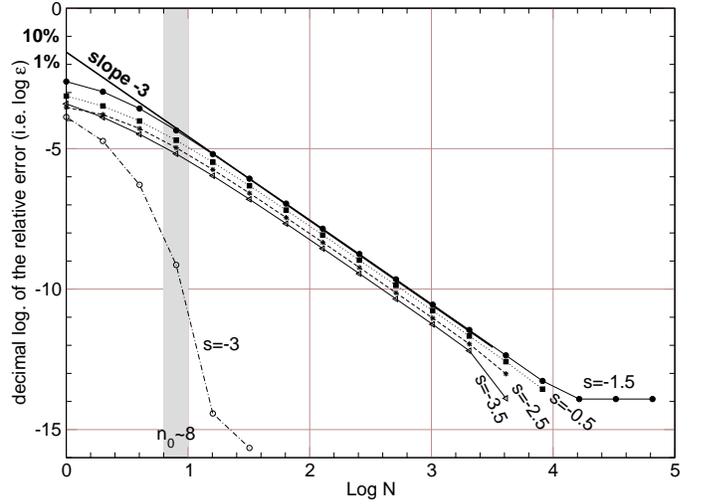}
\caption{Decimal logarithm of the relative error on potential values $\Psi=\Psi_0 + \delta \Psi$ versus the order $N$ of the truncation in Eq.(\ref{eq:psiasympt}) for five power law exponents $s=-3.5$, $-3$, $-2.5$, $-1.5$ and $-0.5$. The disk has inner edge $\ain=0.1$, outer edge $\aout=1$ and the potential is computed at $R=\frac{1}{2}(\ain+\aout)$.}
\label{fig:mpaorders.eps}
\end{figure}

\section{Convergence rate of the series $\delta \Psi = S_\infty(\delta \Psi_n)$ in the case of power law distributions}

\label{sec:crate}

\subsection{A cubic convergence ?}

Figure \ref{fig:mpaorders.eps} shows the rise of the accuracy of potential values $\Psi(R)$ as the order $N$ of the truncation increases for the same disk parameters as already considered and a few exponents $s$. We find that the relative error $\epsilon  \equiv |1- \Psi_{\rm approx.} /\Psi_{\rm ref.}|$ first decreases slowly as $N$ increases. But above a certain rank $n_0 \sim 8-10$, this error approximately scales as $1/N^3$, before saturation near the computer precision, for $N \gtrsim 10^4$ terms. We have checked that this conclusion holds for a wide variety of configurations and radii $R$. Convergence appears to be {\it cubic}. With this approach, we can easily select the lowest order corresponding to a given level of accuracy. From Fig. \ref{fig:mpaorders.eps}, we have roughly:
\begin{equation}
\log \epsilon \approx - 3 \log N  -2 \pm 0.5,
\end{equation}
for $N \gtrsim n_0$, or $N \approx \left( 100 \epsilon \right)^{-1/3}$ within about $0.17 \%$.

There are two remarkable cases. For $s=0$, the potential is fully analytical: $\Psi=\Psi_0$ (i.e. $\delta \Psi = 0$). For $s=-3$, the relative accuracy rises with $N$ much more drastically than for others exponents and reaches the maximum allowed by the computer for $N \gtrsim n_0$ (see below).

\begin{figure}
\includegraphics[width=8.9cm]{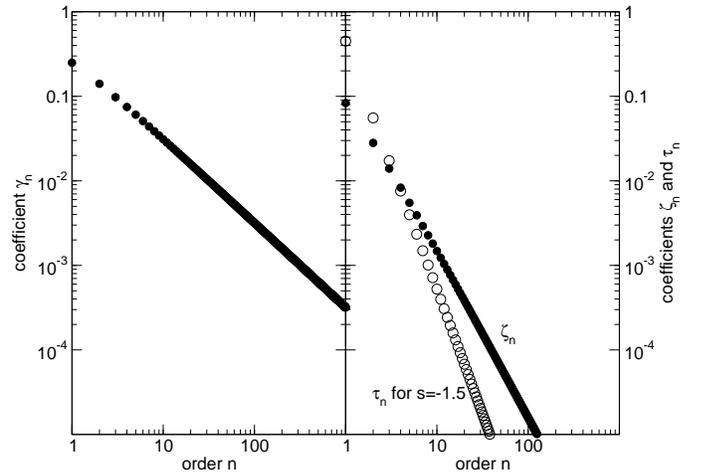}
\caption{{\it Left}: $\gamma_n$ versus $n$. {\it Right} $\zeta_n$ and $\tau_n$ (for $s=-1.5$) versus $n$.}
\label{fig:gamman.eps}
\end{figure}

\begin{figure}
\includegraphics[width=8.9cm]{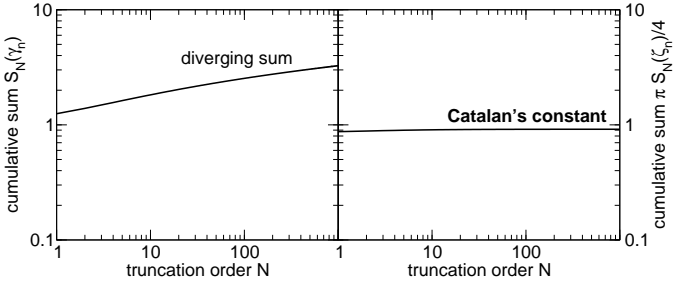}
\caption{{\it Left:} series $S_N(\gamma_n)$ versus the truncation order $N$ which diverges as $N \rightarrow \infty$. {\it Right:}  series $\pi S_N(\zeta_n)/4$ versus the truncation order $N$ which converges towards Catalan's constant (see also Fig. \ref{fig:gamman.eps}).}
\label{fig:sumgamman.eps}
\end{figure}

\subsection{Proof}

The expansion of $\elik(m)$ through Eq. (\ref{eq:kseries}) converges slowly as $m$ approaches unity, and even diverges for $m=1$, although $\gamma_n$ decreases rapidly as $n$ increases (for large $n$, $\gamma_n \sim 1/n$). This is shown in Figs. \ref{fig:gamman.eps} and \ref{fig:sumgamman.eps}. Note that the divergence of $\elik$ as $m \rightarrow 1$ is the main reason why the determination of the potential inside sources by direct integration is generally not the appropriate way. What about the convergence of $\delta \Psi$ ? Since $\delta \Psi_n$ comes from the integration of $\elik$ over its modulus, $S_N(\delta \Psi_n)$ is made of terms of the form $\frac{\gamma_n}{2n+1} \equiv \zeta_n$ which decrease faster with $n$ than terms $\gamma_n$ do. The series $\delta \Psi$ is then expected to converge (see the Appendix \ref{sec:convercatalan}). Figure \ref{fig:gamman.eps} displays the terms $\zeta_n$ versus $n$, and the series $S_N (\zeta_n)$ is plotted versus $N$ in Fig. \ref{fig:sumgamman.eps}. 

 When the series $\delta \Psi$ is truncated at order $N$, the relative error $\epsilon(N)$ on the potential value is:
\begin{equation}
\epsilon(N) = \left| \frac{\Psi_{\rm approx.} - \Psi}{\Psi} \right|.
\end{equation}
From Eqs.(\ref{eq:psiapprox}) and (\ref{eq:psiasympt}), this expression can be re-written in the following form:
\begin{equation}
\epsilon(N) = \left| \frac{S_N(\delta \Psi_n) - S_\infty(\delta \Psi_n)}{\Psi}\right|.
\end{equation}

The quantities $\uin$ and $\vout$ being always smaller than unity, their contribution in Eq.(\ref{eq:deltapsinplaws})  | an exponential drop | is expected to vanish rapidly as $n$ exceeds a certain rank. This transition occurs at $N~\approx~n_0$; it is clearly visible in Fig. \ref{fig:mpaorders.eps} where $n_0 \sim 8$. Thus, for $N \gtrsim n_0$, the leading terms in $\delta \Psi$ form a series $S_N(\tau_n)$, with:
\begin{equation}
\begin{aligned}
\tau_n &= \gamma_n \left[ \frac{1}{2(n+1)} - \frac{1}{2(n+1)+s} +\frac{1}{2n-1}\right.\\
& \qquad \qquad \left. - \frac{1}{2n-1-s}\right],
\end{aligned}
\label{eq:taun}
\end{equation}
which can easily be rearranged:
\begin{equation}
\tau_n = - \gamma_n \frac{s(s+3)(4n+1)}{(2n+2)(2n+2+s)(2n-1)(2n-1-s)}.
\end{equation}
 For large values of $n$, we have:
\begin{equation}
\tau_n \approx -  s(s+3) \frac{\gamma_n}{4 n^3} \left(1 - \frac{3}{4n} + \dots \right).
\label{eq:taunasymp}
\end{equation}
and so, $\tau_n$ behaves like $n^{-4}$, asymptotically. The coefficient $\tau_n$ is displayed versus $n$ for $s=-1.5$ in Fig. \ref{fig:gamman.eps}. Then, as soon as $N \gg n_0$, the relative error is 
\begin{equation}
\epsilon(N) =  \frac{2 \pi G \Sigma_0 R}{\Psi} \left(\sum_{n_0}^N{\tau_n} - \sum_{n_0}^\infty{\tau_n} \right)
\end{equation}
Using Eq.(\ref{eq:taunasymp}), we have:
\begin{equation}
\begin{aligned}
\epsilon(N) & \approx  \frac{2 \pi G \Sigma_0 R}{\Psi} \int_\infty^N{\tau_n dn}\\
            & \approx  - \frac{2 \pi G \Sigma_0 R s(s+3)}{\Psi} \int_\infty^N{\frac{\gamma_n}{4n^3} dn}
\end{aligned}
\end{equation}
Since $\gamma_n$ scales as $1/n$ for large $n$ (see also Fig.(\ref{fig:gamman.eps})), we have:
\begin{equation}
\epsilon(N) \propto - \frac{\Sigma_0 R s(s+3)}{\Psi} \frac{1}{N^3},
\end{equation}
as found ``experimentally''. The convergence rate of this series is therefore {\it cubic}. This result holds irrespective of the radius $R$ (except at the disk edges\footnote{A similar calculus shows that convergence is only {\it quadratic}.}) and whatever $s$, except in the case of the homogeneous disk (i.e. for $s=0$) and for $s=-3$ as well. In this latter case, the convergence rate is not limited by the terms $\tau_n$, but by the series asssociated with $\uin$ and $\vout$. For $s=-3$, we have:
\begin{equation}
\begin{aligned}
\delta \Psi_n & = 2 \pi G \Sigma_0 R \gamma_n \left[\frac{\vout^{2(n+1)}-\uin^{2(n+1)}}{2(n+1)}-\frac{\vout^{2n-1} - \uin^{2n-1}}{2n-1}  \right],
\end{aligned}
\end{equation}
and so the convergence of $\delta \Psi$ is {\it exponential} (see Fig. \ref{fig:mpaorders.eps}), except at the inner and outer edges (see Appendix \ref{sec:conv-3}).

All numerical experiments have succesfully illustrated these results. This is a major issue, in particular because classical multipole expansions converge very slowly inside sources.

\section{Summary and concluding remarks}

In this paper, we have reported a method to determine exact and approximate potential/surface density pairs in flat finite size axially symmetrical disks, in the continuity of the numerical splitting method  described in Pierens \& Hur\'e (2005). At a given radius $R$, each pair is the sum of two components:
\begin{itemize}
\item a ``homogeneous'' pair $(\Psi_0,\Sigma_0)$ where $\Sigma_0= \Sigma(R)$ is the local value of the surface density and $\Psi_0$ is given by Eq.(\ref{eq:psi0}),
\item a ``residual'' pair $(\delta \Psi, \delta \Sigma)$, where $\delta \Psi$ is an infinite series (see Eqs.(\ref{eq:psiinf}) and(\ref{eq:deltapsin})) and $\delta \Sigma(a) = \Sigma(a) - \Sigma_0$.
\end{itemize}
In general, the infinite series $\delta \Psi$ cannot be put into a closed form expression. However, in contrast to the multipole expansions commonly met in potential theory, {\it this series has a cubic convergence rate inside the source}. This major result is due to the splitting method which is equivalent to a convergence acceleration technique (in the mathematical sense). It means that very good appro\-ximations for the potential can be obtained by low order truncations of the series. Pairs can be constructed only for a certain class of the surface density profile $\Sigma(a)$: any function $\Sigma(a)$ whose product by a power law of the radius $a$ is analytically integrable does lead to a potential/surface density pair. This is the case of power laws. It would be interesting to list all functions satisfying this criterion.

We have fully considered power law surface densities. Such profiles are particularly attractive for two main reasons: i) power laws underly many (if not most) astrophysical disk models, and ii) power laws may serve as a basis set to construct more complicated surface density profiles. This new approach along with associated computing tools\footnote{An easy-to-use Fortran 90 package nammed {\tt APPLawD} (an acronym for ``Accurate Potential for Power Law Disks'') is available upon request to the authors. {\tt APPLawD} can be downloaded at the following address:\\ {\tt www.obs.u-bordeaux1.fr/radio/JMHure/intro2applawd.html}}
 should enable a much better representation of the gravitational potentials and forces in flat disks.

\begin{acknowledgements}
We thank J. Braine for many fruitful comments. We are especially grateful to the referee L. Ciotti whose suggestions and criticisms have allowed significantly improvement to the paper.
\end{acknowledgements}

\bibliographystyle{aa}

\appendix

\section{Special cases for $\delta \Psi_0$}
\label{sec:sp0}

For $s=-1$ in Eq.(\ref{eq:sigmaplaw}), then Eq.(\ref{eq:deltapsi0}) reads
\begin{equation}
\delta \Psi_0= 2 \pi G \Sigma_0 R\left\{ \left[\frac{u^2}{2}- u\right]_{\uin}^1 + \left[ \ln v + \frac{1}{v} \right]_1^{\vout} \right\}
\end{equation}
For $s=-2$ in Eq.(\ref{eq:sigmaplaw}), then Eq.(\ref{eq:deltapsi0}) reads
\begin{equation}
\delta \Psi_0= 2 \pi G \Sigma_0 R\left\{ \left[\frac{u^2}{2}-\ln u\right]_{\uin}^1 + \left[ v + \frac{1}{v} \right]_1^{\vout} \right\}
\end{equation}

\section{Special cases for $\delta \Psi_1$}
\label{sec:sp1}

For $s=-4$ in Eq.(\ref{eq:sigmaplaw}), then Eq.(\ref{eq:deltapsi1}) reads
\begin{equation}
\delta \Psi_1= \frac{\pi}{2} G \Sigma_0 R\left\{ \left[\frac{u^4}{4}-\ln u\right]_{\uin}^1 + \left[ \frac{v^5}{5} - v \right]_1^{\vout} \right\}
\end{equation}
For $s=+1$ in Eq.(\ref{eq:sigmaplaw}), then Eq.(\ref{eq:deltapsi1}) reads
\begin{equation}
\delta \Psi_1= \frac{\pi}{2} G \Sigma_0 R\left\{\left[\frac{u^4}{4}-\frac{u^5}{5}\right]_{\uin}^1 + \left[ \ln v - v \right]_1^{\vout} \right\}
\end{equation}

\section{Special cases for $\delta \Psi_n$}
\label{sec:spn}

For $s+2n+2=0$ in Eq.(\ref{eq:sigmaplaw}), then $s-2n+1 \ne 0$ and Eq.(\ref{eq:deltapsin}) reads
\begin{flalign}
\delta \Psi_n &= 2 \pi G \Sigma_0 R \gamma_n \left\{ \left[\frac{u^{2(n+1)}}{2(n+1)}-\ln u\right]_{\uin}^1 \right. \\
                 &\qquad \qquad \qquad \qquad \left. + \left[ \frac{v^{2n-s-1}}{2n-s-1} - \frac{v^{2n-1}}{2n-1} \right]_1^{\vout} \right\}
\nonumber
\end{flalign}

For $s+2n+2\ne0$ and $s-2n+1=0$ in Eq.(\ref{eq:sigmaplaw}), then Eq.(\ref{eq:deltapsin}) reads
\begin{flalign}
\delta \Psi_n &= 2 \pi G \Sigma_0 R \gamma_n \left\{ \left[\frac{u^{2(n+1)}}{2(n+1)}-\frac{u^{s+2(n+1)}}{s+2(n+1)}\right]_{\uin}^1 \right. \\
                 &\qquad \qquad\qquad \qquad \qquad \qquad \left. + \left[ \ln v -  \frac{v^{2n-1}}{2n-1} \right]_1^{\vout} \right\}
\nonumber
\end{flalign}

\section{Convergence of $S_\infty (\zeta_n)$}
\label{sec:convercatalan}

From \cite[e.g.][]{gradryz65}, we see that
\begin{equation}
\begin{aligned}
 S_\infty (\zeta_n) &= \sum_{n=0}^\infty{\frac{\gamma_n}{2n+1}},\\
 &= \sum_{n=0}^\infty{\int_0^1{\gamma_n m^{2n} dm}},\\
 &= \int_0^1{\sum_{n=0}^\infty{\gamma_n m^{2n} dm}},\\
 &= \frac{2}{\pi} \int_0^1{\elik(m)dm},\\
 & \equiv \frac{4C}{\pi},
\end{aligned}
\end{equation}
where $C$ is Catalan's constant. 

\section{Convergence rate for $s=-3$ at the disk edges}
\label{sec:conv-3}

At the disk outer edge where $\vout=1$, we have:
\begin{equation}
\delta \Psi_n = 2 \pi G \Sigma_0 R \gamma_n \left[\frac{1-\Delta^{2(n+1)}}{2(n+1)}-\frac{1 - \Delta^{2n-1}}{2n-1}  \right],
\end{equation}
where $\Delta=\ain/\aout$ is the disk axis ratio. Since $\Delta < 1$, the leading terms in $\delta \Psi$ then form a series $S_N(\tau'_n)$, with:
\begin{equation}
\begin{aligned}
\tau'_n &= \gamma_n \left[ \frac{1}{2(n+1)} - \frac{1}{2n-1} \right],\\
        & = -\gamma_n \frac{3}{2(n+1)(2n-1)}.\\
\end{aligned}
\label{eq:tauprimn}
\end{equation}
For large $n$, we have $\tau'_n \sim 1/n^3$, and so $\epsilon(N) \propto 1/N^2$. The convergence is therefore {\it quadratic} (as at the disk inner edge).

\end{document}